\documentclass[nopacs,twocolumn,prl,aps,color,superscriptaddress]{revtex4}
\usepackage{graphicx}  
\usepackage{bm}        
\usepackage{amssymb}   

\begin{document}

\title{Robust Magnetoelectric Effect in Decorated Graphene/In$_{2}$Se$_{3}$ Heterostructure}

\author{Jing Shang}
\affiliation{School of Mechanical, Medical and Process Engineering, Queensland University of Technology, Brisbane, QLD 4001, Australia}
\author{Xiao Tang}
\affiliation{School of Mechanical, Medical and Process Engineering, Queensland University of Technology, Brisbane, QLD 4001, Australia}
\author{Yuantong Gu}
\affiliation{School of Mechanical, Medical and Process Engineering, Queensland University of Technology, Brisbane, QLD 4001, Australia}
\author{Arkady V. Krasheninnikov}
\affiliation{Institute of Ion Beam Physics and Materials Research, Helmholtz-Zentrum Dresden-Rossendorf, 01328 Dresden, Germany}
\author{Silvia Picozzi}
\email{silvia.Picozzi@spin.cnr.it}
\affiliation{Consiglio Nazionale Delle Ricerche, Istituto SPIN, UOS l'Aquila, Sede di Lavoro CNR-SPIN C/o Universitá G. d'Annunzio, Chieti, 66100, Italy}
\author{Changfeng Chen}
\email{chen@physics.unlv.edu}
\affiliation{Department of Physics and Astronomy, University of Nevada, Las Vegas, Nevada 89154, United States}
\author{Liangzhi Kou}
\email{liangzhi.kou@qut.edu.au}
\affiliation{School of Mechanical, Medical and Process Engineering, Queensland University of Technology, Brisbane, QLD 4001, Australia}

\date{\today}

\begin{abstract}
Magnetoelectric effect is a fundamental physics phenomenon that synergizes electric and magnetic degrees of freedom to generate distinct material responses like 
electrically tuned magnetism, which serves as a key foundation of the emerging field of spintronics. Here, we show by first-principles studies that ferroelectric 
(FE) polarization of an In$_{2}$Se$_{3}$ monolayer can modulate the magnetism of an adjacent transition-metal (TM) decorated graphene layer via an FE induced 
electronic transition. The TM nonbonding $d$-orbital shifts downward and hybridizes with carbon $p$ states near the Fermi level, suppressing the magnetic moment,
under one FE polarization, but on reversed FE polarization this TM $d$-orbital moves upward, restoring the original magnetic moment. This finding of robust magnetoelectric
effect in TM decorated graphene/In$_2$Se$_3$ heterostructure offers powerful insights and a promising avenue for experimental exploration of FE controlled magnetism 
in 2D materials.
\end{abstract}
\maketitle

Low-dimensional magnetic materials provide excellent platforms for spintronics that use electron spin rather than charge as the information carrier, and recent years
have seen tremendous developments in this emerging field that promises equipments with higher storage density and lower energy consumption {\cite{ref1,ref2,ref3}}.
Prototypical spintronic devices have been proposed and demonstrated \cite{ref4,ref5,ref6}, but further advances of the field have been hindered by a lack of suitable low-dimensional magnetic materials and adequate means for effectively tuning their magnetic behaviors. Recent studies have reported synthesis of atomically thin
2D magnets, such as CrI$_{3}$ \cite{ref7}, CrGeTe$_{3}$ \cite{ref8}, Fe$_{3}$GeTe$_{2}$ \cite{ref9} and Fe$_{2}$O$_{3}$ \cite{ref10}; but these materials have low
Curie temperatures ($\sim$45 K) that limit the scope of their viability. Meanwhile, graphene as the first discovered truly 2D material \cite{ref11} has been
explored for its magnetic properties \cite{ref12}, and magnetism was observed in transition-metal (TM) decorated graphene \cite{ref13,ref14,ref15,ref16,ref17,ref18,ref19}
induced by the hybridization of carbon \emph{p} and TM \emph{d} orbitals \cite{ref20,ref21}. Specially notable is the synthesis of suspended TM decorated single vacancy
graphene (TM@SVG) monolayer \cite{ref22,ref23} with considerably improved stability of well dispersed TM adsorption.

Effective control of magnetism in low-dimensional materials by reliable and convenient means is an essential requirement in spintronics. Among various strategies,
electric-field controlled magnetism is regarded as the most promising \cite{ref24,ref25,ref26,ref27,ref28}. There are mainly three material-class specific working
mechanisms \cite{ref26}: (i) magnetic exchange modulation by electrically tuneable carrier concentration in magnetic semiconductors, (ii) changing coercivity or
magnetic anisotropy by shifting the Fermi level in magnetic metals, and (iii) magnetic response to electric tuning via magnetoelectric coupling in multiferroics.
Major challenges remain, however, in identifying suitable low-dimensional, mainly 2D, materials with robust magnetism that is responsive to electric-field tuning.
A major development on this front is the recently reported synthesis of In$_{2}$Se$_{3}$ monolayer that has proved to be a versatile platform for designing and
implementing 2D ferroelectric (FE) based nonvolatile memory devices \cite{ref29,ref30} and also for facilitating diverse chemical and physical processes, such as
photocatalytic water splitting \cite{ref31} and magnetic anisotropy modulations \cite{ref32}.

In this Letter, we present computational evidence for robust magnetoelectric effect in viable TM@SVG/In$_{2}$Se$_{3}$ heterostructures, allowing for effective electric
tuning of magnetism. First-principles calculations show that magnetic moments derived from select TM atoms adsorbed on SVG (V, Cr, Mn@SVG) are highly sensitive to,
and therefore effectively controllable by, the switch of FE polarization of the In$_{2}$Se$_{3}$ monolayer. This intriguing phenomenon stems from an FE-induced electronic
transition that shifts the originally unoccupied nonbonding TM-$d$ states toward the Fermi level to hybridize with the carbon-derived states, thereby largely suppressing
the magnetic moment, while a reversal of FE polarization restores the original magnetic state. This electric modulation of magnetism relies on the strong characters of FE polarization of In$_2$Se$_3$ monolayer, which exerts FE-sensitive electrostatic potentials and associated electron transfers across the interface in the heterostructure.
These findings unveil a distinct mechanism for compelling FE controlled magnetism, making TM@SVG/In$_{2}$Se$_{3}$ a promising platform for exploring new magnetoelectric
effects in advanced spintronics materials research.

\begin{figure}
\includegraphics[width=8.5cm]{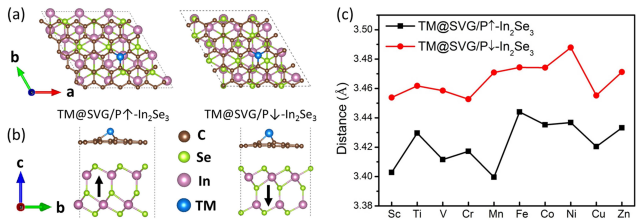}
\caption{\label{fig1} Atomic configurations in (a) top and (b) side view of TM@SVG/In$_{2}$Se$_{3}$ heterostructure with the ferroelectric polarizations in two characteristic
directions, defined as up and down, indicated by the black arrows in (b). Results in (c) show the distances between the TM@SVG with various TM atoms and In$_2$Se$_3$ monolayer in two distinct polarizations.}
\end{figure}

Spin-polarized electronic band-structure and total energy calculations have been performed using the Vienna Ab-initio Simulation Package (VASP) \cite{ref33,ref34}.
Computational details are provided in the Supplemental Material \cite{ref35}. The TM@SVG/In$_{2}$Se$_{3}$ heterostructure is constructed by matching a 5$\times$5$\times$1
supercell of SVG with a 3$\times$3$\times$1 supercell of In$_{2}$Se$_{3}$ monolayer, as shown in Fig.~\ref{fig1}. The lattice mismatch in this
structural model is 0.2$\%$, which has only minimal effects on the computed total energies and structural and electronic properties. The In$_{2}$Se$_{3}$ monolayer has two
distinct faces in contact with graphene, with the FE polarization either pointing upward (P$\uparrow$) or downward (P$\downarrow$), as defined in Fig.~\ref{fig1}(b).

We first examined TM decorated graphene on In$_{2}$Se$_{3}$ (TM@graphene/In$_{2}$Se$_{3}$), where the TM atom is adsorbed at the energetically most preferred hexagonal hollow site (see Fig. S1(a-b) \cite{ref35}), and the FE polarization switch of In$_{2}$Se$_{3}$ monolayer produces obvious effects on the total magnetic moments of the system [Fig. S1(d)]. For Cr or Mn, the magnetic moments are 0$\mu_B$ at TM@graphene/P$\uparrow$-In$_{2}$Se$_{3}$, but become $\sim$4.5$\mu_B$ at TM@graphene/P$\downarrow$-In$_{2}$Se$_{3}$ (these magnetic moments are reduced slightly by on-site Coulomb interaction, see Fig. S1(c) \cite{ref35}). These results demonstrate significant tunability of FE controlled magnetism. However, TM atoms adsorbed on graphene have low migration energy barriers (0.2$\sim$0.8 eV \cite{ref21}) that create a strong tendency for TM atoms to migrate and
form metal clusters, which is detrimental to the stability and functionality of TM@graphene/In$_{2}$Se$_{3}$. A promising resolution to this issue is offered by the recently synthesized suspended TM@SVG layered structure where the TM atom bonds strongly to the otherwise under-coordinated carbon atoms near the vacancy site in the graphene layer,
thus greatly enhancing the stability of the dispersed TM atom configuration and making TM@SVG a viable layered magnetic structure, which can be used to form functional heterostructgures with In$_{2}$Se$_{3}$ monolayer. Our calculations show that TM atoms in TM@SVG take out-of-layer equilibrium positions due to its larger radius compared to
carbon, which is consistent with previous reports \cite{ref20,ref21}. Meanwhile, the interlayer distance between graphene and In$_{2}$Se$_{3}$ is sensitive to the FE polarization of the latter [Fig.~\ref{fig1}(c)], which reflects the distinct nature of the FE-dependent interlayer interaction and has profound effects on key properties of the heterostruture as will be discussed below.

\begin{figure}
\includegraphics[width=8.5cm]{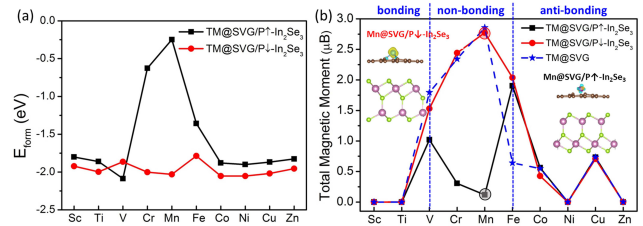}
\caption{\label{fig2} (a) Formation energy of TM@SVG for various TM atoms and In$_{2}$Se$_{3}$ monolayer in up (P$\uparrow$) or down (P$\downarrow$) FE polarization.
(b) Total magnetic moment M ($\mu_B$) of TM@SVG/In$_{2}$Se$_{3}$ heterostructures. Insets show the spin density of Mn@SVG/In$_{2}$Se$_{3}$ with both FE polarizations in In$_{2}$Se$_{3}$, where the isosurface is set to 0.03 eV/{\AA}$^{3}$.}
\end{figure}

To assess the stability of TM@SVG/In$_2$Se$_3$ heterostructures and the influence of the FE polarization in the In$_2$Se$_3$ monolayer, we have calculated formation energy E$_{form}$ = E$_{total}$-E$_{TM@SVG}$-E$_{\text{In$_{2}$Se$_{3}$}}$, where E$_{total}$, E$_{TM@SVG}$, and E$_{\text{In$_{2}$Se$_{3}$}}$ are, respectively, the energy of the heterostructure, freestanding TM@SVG, and In$_{2}$Se$_{3}$ monolayer. This quantity measures the interaction between the TM@SVG and In$_{2}$Se$_{3}$ layers, and the calculated results for all the studied TM cases are negative as shown in Fig.~\ref{fig2}(a), indicating that these heterostructures are viable. It is noted that formation energies of TM@SVG/P$\downarrow$-In$_{2}$Se$_{3}$ are generally larger than those with P$\uparrow$ FE polarization, except for V@SVG/In$_{2}$Se$_{3}$. For most TM@SVG cases, differences in formation energy with P$\downarrow$ and P$\uparrow$ polarizations are small, around 0.1 eV, but two outstanding cases, (Cr,Mn)@SVG/In$_{2}$Se$_{3}$, exhibit considerably larger differences in polarization dependent formation energy, up to 1.5$\sim$2 eV, indicting strong contrasting effects by the opposite FE states in the In$_2$Se$_3$ monolayer.

\begin{figure}
\includegraphics[width=8.5cm]{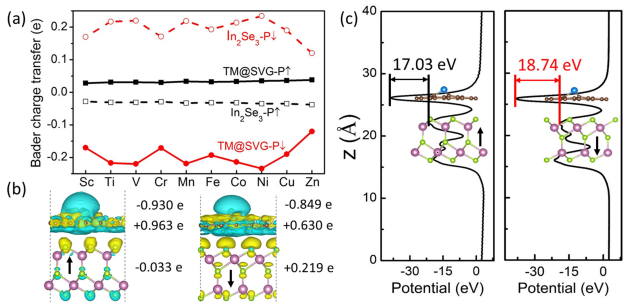}
\caption{\label{fig3} (a) Electron transfer based on a Bader charge analysis between TM@SVG and In$_{2}$Se$_{3}$ layers in TM@SVG/In$_{2}$Se$_{3}$ heterostructures. (b) The differential charge density of Mn@SVG/In$_{2}$Se$_{3}$ heterostructures with P$\uparrow$ and P$\downarrow$ FE polarization in the In$_{2}$Se$_{3}$ layer. The isosurface value is set to 0.0002 eV/{\AA}$^{3}$. (c) The electrostatic potential profile for Mn@SVG/In$_{2}$Se$_{3}$ and the potential differences between graphene and contacting In atoms for the two polarizations.}
\end{figure}

We now evaluate the FE polarization dependence of magnetic properties of TM@SVG/In$_{2}$Se$_{3}$. The total magnetic moments with the In$_2$Se$_3$ layer in P$\uparrow$ or
P$\downarrow$ FE polarizations are displayed in Fig.~\ref{fig2}(b), compared with the results for freestanding TM@SVG. It has been proposed \cite{ref20} that these TM@SVG systems mainly fall into three categories: bonding (Sc, Ti), nonbonding (V, Cr, Mn) and antibonding (Co, Ni, Cu, Zn), while Fe represents a special case, sitting between nonbonding and antibonding regimes and exhibiting complex magnetic behaviors. Our calculated results shown in Fig.~\ref{fig2}(b) indicate that (i) In$_{2}$Se$_{3}$ in P$\downarrow$ polarization has little influence on the magnetic moments of the TM@SVG studied here, with the only exception for Fe@SVG, all of which nearly coincide with result of the freestanding TM@SVG layer; (ii) for the bonding cases, the TM \emph{d} orbitals are fully occupied, rendering zero magnetic moment, independent of the FE polarization of In$_2$Se$_3$, and for the antibonding cases, the results for TM@SVG/P$\downarrow$-In$_{2}$Se$_{3}$, TM@SVG/P$\uparrow$-In$_{2}$Se$_{3}$ and freestanding TM@SVG all stay very close; (iii) the nonbonding cases of V, Cr and Mn@SVG/In$_{2}$Se$_{3}$ possess magnetic moments that are sensitively dependent on the FE polarization of In$_2$Se$_3$. For example, the magnetic moment is 2.9$\mu_B$ at Mn@SVG/P$\downarrow$-In$_{2}$Se$_{3}$, but only 0.12$\mu_B$ at Mn@SVG/P$\uparrow$-In$_{2}$Se$_{3}$. To probe the origin of these sharply contrasting behaviors, we have artificially tuned the distance between Mn@SVG and In$_{2}$Se$_{3}$ layers to test the influence of the interlayer interaction, and the results [see Fig. S2 \cite{ref35}] show that the magnetic moment of Mn@SVG/P$\uparrow$-In$_{2}$Se$_{3}$ increases with the rising interlayer distance, suggesting strong interlayer effect on the magnetism on Mn@SVG layer, and the value approaches that of the freestanding TM@SVG at large interlayer distances when the effect of the In$_2$Se$_3$ layer on the TM@SVG diminishes. Meanwhile, the magnetic moment of Mn@SVG/P$\downarrow$-In$_{2}$Se$_{3}$ remains largely unchanged throughout this process, indicating little influence by the interlayer interaction.
The FE polarization dependent magnetism is also seen in spatial spin distributions, where a larger amount of unpaired electrons gather around the Mn site in the case of Mn@SVG/P$\downarrow$-In$_{2}$Se$_{3}$, while a much smaller spin density appears in the case of Mn@SVG/P$\uparrow$-In$_{2}$Se$_{3}$ [see insets of Fig.~\ref{fig2}(b)].

To check the dependence of the calculated interlayer interaction and associated magnetic behaviors on the choice of the vdW potential, we have performed additional
calculations using the Tkatchenko-Scheffler (TS) and DFT-D3 method with Becke-Jonson damping methods. For the exemplary case of Mn@SVG/In$_{2}$Se$_{3}$, we find that
although the magnitude of magnetic moments do vary quantitatively with the choice of potentials, the overall behaviors remain unchanged (for details see the results
presented in Table S1 \cite{ref35}).

\begin{figure}
\includegraphics[width=8.5cm]{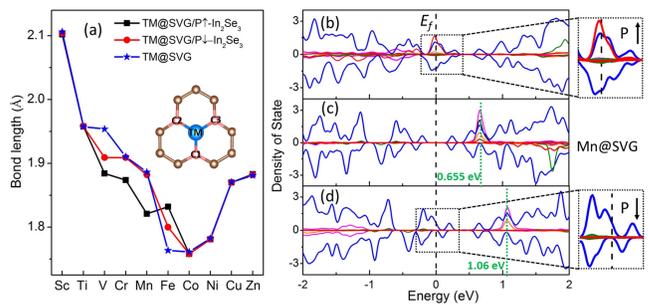}
\caption{\label{fig4} (a) The bond length between the TM atom and the nearest-neighbor carbon atoms (inset shows the structural configuration) in the TM@SVG/In$_2$Se$_3$ heterostructure with distinct FE polarizations and in the freestanding TM@SVG. (b-d) Spin polarized (positive/negative values for up/down spin) electronic density of states;
blue and red lines represent contributions from the carbon $p_z$ and Mn $d_{x^2-y^2}$ orbitals.}
\end{figure}

The sharply contrasting behaviors of the formation energy and magnetic responses of the TM@SVG-In$_2$Se$_3$ heterostructure presented above can be understood in terms of the interfacial charge transfer and the resulting shift of the TM \emph{d} orbital under distinct FE polarizations of the In$_2$Se$_3$ monolayer. The results of a Bader charge analysis [Fig.~\ref{fig3}(a)] indicate that opposite FE polarizations have significantly distinct influence on the electron transfer between the TM@SVG and In$_{2}$Se$_{3}$ layers. When In$_{2}$Se$_{3}$ is in the P$\uparrow$ polarization, only 0.03\emph{e} is transferred to TM@SVG from In$_{2}$Se$_{3}$, which acts as an electron donator. In sharp contrast, In$_{2}$Se$_{3}$ becomes an electron acceptor to receive a much larger amount ($\sim$0.2\emph{e}) of charge transfer from the TM@SVG layer when the polarization is reversed to P$\downarrow$.  This drastically increased electron transfer leads to much stronger interfacial interaction, as reflected in the greatly enhanced formation energy [Fig.~\ref{fig2}(a)]. For an intuitive understanding, we take Mn@SVG/In$_{2}$Se$_{3}$ as an example to analyze the spatial charge density difference $\Delta \rho$ = $\rho_{total}$ - $\rho_{In_2Se_3}$ - $\rho_{SVG}$ - $\rho_{Mn}$. The results [Fig.~\ref{fig3}(b)] show that in P$\uparrow$-In$_{2}$Se$_{3}$ electron transfer occurs mostly from the Mn atom to SVG with only a minimal amount from the In$_2$Se$_3$ monolayer, whereas in P$\downarrow$-In$_{2}$Se$_{3}$ there is a considerable amount of electron transfer from the Mn atom to the In$_{2}$Se$_{3}$ monolayer through the contacting SVG layer. This result can be ascribed to the larger electrostatic potential difference (18.74 eV versus 17.03 eV) at the interfaces of the heterostructure as shown in Fig.~\ref{fig3}(c), where the surface of the In$_2$Se$_3$ monolayer with lower electrostatic potential is in close contact with the Mn@SVG layer.

The FE controlled magnetic behaviors [Fig.~\ref{fig2}(b)] can be further elucidated by examining pertinent chemical bonding changes and the spin polarized density of states of the adsorbed TM atom under different polarizations of the In$_{2}$Se$_{3}$ monolayer. Since most of the magnetic moment is contributed by the TM and surrounding carbon atoms,
it is instructive to analyze the TM-C bond length and the associated band shifts. We show in  Fig.~\ref{fig4}(a) the calculated TM-C bond length, and the results indicate that the TM-C bond length first follows a descending trend in going from 2.10 {\AA} (Sc) to 1.75 {\AA} (Co), and then moves upward and increases to 1.90 {\AA} (Zn);
the same trend is seen in freestanding TM@SVG \cite{ref35}. For most cases, the bond lengths between the adsorbed TM atom and the three nearest-neighbor carbon atoms remain nearly unchanged under the switch of FE polarization, so the C$_{3v}$ symmetry at the TM-adsorbed carbon vacancy site in SVG is preserved during the FE switch. Notable differences, however, exist in the TM-C bond length for the cases involving the nonbonding states (V, Cr, Mn). This pattern in bond-length variations corroborates with the behaviors of the magnetic modulation by FE polarization [Fig.~\ref{fig2}(b)]. These shorter TM-C bonds enhance the hybridization between the TM-$d$ and carbon-$p$ states,
thereby asserting stronger influence on the magnetic moment of the TM@SVG-In$_2$Se$_3$ heterostructure.

For a more in-depth assessment of the driving mechanism for the magnetic responses, we have examined the partial density of state (PDOS) of the Mn-\emph{d} orbital and its hybridization with the \emph{p} orbitals of the surrounding nearest-neighbor carbon atoms. Calculated spin-polarized PDOS for Mn@SVG-In$_2$Se$_3$ in both polarizations are
shown in Fig.~\ref{fig4}(c), together with the results for freestanding Mn@SVG for comparison. It is seen that in Mn@SVG/P$\uparrow$-In$_{2}$Se$_{3}$) the shortening of the
Mn-C bonds leads to a large downward shift of the Mn-\emph{d} orbital, which is located 0.655 eV above the Fermi level in freestanding Mn@SVG, to the vicinity of the Fermi
level. As a result, the originally unoccupied nonbonding Mn-\emph{d} state is now partially occupied and hybridizes with the nearest C-\emph{p} states near the Fermi level.
This large shifting of the Mn-$d$ state considerably reduces the spin spitting and, therefore the associated magnetic moment. Similar mechanisms have been invoked in previous
studies \cite{ref36,ref37} to explain the relationship between carbon vacancies and the induced magnetism. In sharp contrast, after the FE polarization of the heterostructure 
is switched to the reversed direction, the Mn-$d$ state in Mn@SVG/P$\downarrow$-In$_{2}$Se$_{3}$) moves upward in energy from its original position to be further away, located
at 1.06 eV above the Fermi level, accompanied by a slight shift of the C-\emph{p} states toward the Fermi level. Under this circumstance, there is no hybridization between the Mn-\emph{d} and C-\emph{p} orbitals, and the pertinent nonbonding Mn-$d$ state remains unoccupied, thus leaving the magnetic moment of the heterostruture unaffected in P$\downarrow$-In$_{2}$Se$_{3}$, despite the resulted increase in the spin splitting of the Mn-$d$ states. This mechanism also renders similar phenomena of FE polarization modulation in Cr@SVG/In$_{2}$Se$_{3}$ (see Fig. S3 \cite{ref35}). In the P$\uparrow$ FE polarization of the In$_2$Se$_3$ monolayer, the the Cr-\emph{d} state shifts downward toward the Fermi level and hybridizes with the C-\emph{p} states, resulting in the diminished total magnetic moment in the heterostructure. In contrast, when the FE polarization is reserved to P$\downarrow$, the nonbonding Cr-$d$ state moves upward and remain unoccupied, thus not affecting the magnetism. These results suggest that the reversible FE polarization of In$_{2}$Se$_{3}$ monolayer can act as an effective control for magnetism of TM@SVG in a heterostructure configuration, which is highly promising for feasible and convenient modulation of 2D magnetism that may facilitate innovative design and implementation in novel spintronic operations.

In summary, we find robust magnetoelectric effect in select TM@SVG/In$_2$Se$_3$ heterostructure based on results from systematic first-principles calculations. Our extensive computational studies show that switching the FE polarization in In$_{2}$Se$_{3}$ monolayer can effectively modulate the magnetic moment in adjacent TM@SVG layer, thereby realizing a long-sought platform for feasible experimental exploration of effective electric control of magnetism. An analysis of the interfacial bonding and charge transfer in the heterostructure reveals sensitive dependence of the charge and spin polarizations that concurrently impact the FE and magnetic order in the system. In particular, the nonbonding TM-$d$ state originally located above the Fermi level in freestanding TM@SVG shifts in opposite directions under the influence of different FE polarizations. In response to P$\downarrow$-In$_2$Se$_3$, the TM-$d$ state moves further away from the Fermi level, enhancing the spin splitting but not altering the magnetic moment, while under P$\uparrow$-In$_2$Se$_3$, the TM-$d$ state drops below the Fermi level, becoming partially occupied and hybridizing with carbon-$p$ states, which considerably reduces the spin splitting of the TM-$d$ states and the associate dominant contributions to the magnetic moment. Since freestanding magnetic TM@SVG and ferroelectric In$_{2}$Se$_{3}$ monolayer structures both have been experimentally synthesized and characterized, it is reasonable to expect that the construction of the heterostructure based on these 2D materials is practically feasible. The intricate working mechanisms unveiled in the present work may help develop other layered material structures in the fast growing family of 2D materials possessing diverse electric and magnetic behaviors that could combine to form heterostructures with pronounced magnetoelectric effect, thus further advancing and expanding the materials basis for the fast growing field of spintronics research and development.

We acknowledge the grants of high-performance computer time from computing facility at the Queensland University of Technology, the Pawsey Supercomputing Centre and Australian National Computational Infrastructure (NCI). L.K. gratefully acknowledges financial support by the ARC Discovery Project (DP190101607).

\
\bibliographystyle{plain}

\begin{thebibliography}{99}

\bibitem{ref1}Y. P. Feng, L. Shen, M. Yang, A. Wang, M. Zeng, Q. Wu, S.
Chintalapati, and C.-R. Chang, WIREs Comput. Mol. Sci. \textbf{7}, e1313
(2017).

\bibitem{ref2}H. Li, S. Ruan, and Y.-J. Zeng, Adv. Mater. \textbf{31},
1900065 (2019).

\bibitem{ref3}A. Fert, Angew. Chem. Int. Ed. \textbf{47}, 5956 (2008).

\bibitem{ref4}A. Hirohata and K. Takanashi, J. Phys. D: Appl. Phys.
\textbf{47}, 193001 (2014).

\bibitem{ref5}H. Yan, \emph{et al.}, Adv. Mater. \textbf{32}, 1905603
(2020).

\bibitem{ref6}S. D. Bader and S. S. P. Parkin, Annu. Rev. Condens. Matter
Phys. \textbf{1}, 71 (2010).

\bibitem{ref7}B. Huang \emph{et al.}, Nature \textbf{546}, 270 (2017).

\bibitem{ref8}C. Gong \emph{et al.}, Nature \textbf{546}, 265 (2017).

\bibitem{ref9}J. Yi, H. Zhuang, Q. Zou, Z. Wu, G. Cao, S. Tang, S. A.
Calder, P. R. C. Kent, D. Mandrus, and Z. Ga, 2D Mater. \textbf{4}, 011005
(2017).

\bibitem{ref10}A. P. Balan \emph{et al.}, Nat Nanotechnol. \textbf{13}, 602
(2018).

\bibitem{ref11}K. S. Novoselov, A. K. Geim, S. V. Morozov, D. Jiang, Y.
Zhang, S. V. Dubonos, I. V. Grigorieva, and A. A. Firsov, Science
\textbf{306}, 666 (2004).

\bibitem{ref12}A. Candini, S. Klyatskaya, M. Ruben, W. Wernsdorfer, and M.
Affronte, Nano Lett. \textbf{11}, 2634 (2011).

\bibitem{ref13}Y. Mao, J. Yuan, and J. Zhong, J. Phys.: Condens. Matter.
\textbf{20}, 115209 (2008).

\bibitem{ref14}R. Sielemann, Y. Kobayashi, Y. Yoshida, H. P. Gunnlaugsson,
and G. Weyer, Phys. Rev. Lett. \textbf{101},137206 (2008).

\bibitem{ref15}O. Cretu, A. V. Krasheninnikov, J. A. Rodríguez-Manzo, L.
Sun, R. M. Nieminen, and F. Banhart, Phys. Rev. Lett. \textbf{105}, 196102
(2010).

\bibitem{ref16}M. Manadé, F. Vińes, and F. Illas, Carbon \textbf{95}, 525
(2015).

\bibitem{ref17}V. Zólyomi, Á. Rusznyák, J. Kürti, and C. J. Lambert, J.
Phys. Chem. C \textbf{114}, 18548 (2010).

\bibitem{ref18}P. Blonski and J. Hafner, J. Phys. Chem,\textbf{134}, 154705
(2011).

\bibitem{ref19}C. Cao, M. Wu, J. Jiang, and H.-P. Cheng, Phys. Rev. B
\textbf{81}, 205424 (2010).

\bibitem{ref20}E. J. G. Santos, A. Ayuela, and D. Sánchez-Portall, New J.
Phys. \textbf{12}, 053012 (2010).

\bibitem{ref21}A. V. Krasheninnikov, P. O. Lehtinen, A. S. Foster, P.
Pyykkö, and R. M. Nieminen, Phys. Rev. Lett. \textbf{102}, 126807 (2009).

\bibitem{ref22}Q. M. Ramasse, R. Zan, U. Bangert, D. W. Boukhvalov, Y.-W.
Son, and K. S. Novoselov, ACS Nano \textbf{6}, 4063 (2012).

\bibitem{ref23}H. Wang,\emph{et al.}, Nano Lett. \textbf{12}, 141 (2012).

\bibitem{ref24}C. C. Homes, \emph{et al.}, Nature \textbf{430}, 539 (2004).

\bibitem{ref25}C. Ederer and N. A. Spaldin, Phys. Rev. B \textbf{74},
020401(R) (2006).

\bibitem{ref26}F. Matsukura, Y. Tokura, and H. Ohno, Nat. Nanotechnol.
\textbf{10}, 209 (2015).

\bibitem{ref27}P. Lou and J. Y. Lee, J. Phys. Chem. C \textbf{113}, 21213
(2009).

\bibitem{ref28}Y.-H. Chu, L. W. Martin, M. B. Holcomb, and R. Ramesh,
Mater. Today \textbf{10}, 16 (2007).

\bibitem{ref29}W. Ding, J. Zhu, Z. Wang, Y. Gao, D. Xiao, Y. Gu, Z. Zhang,
and W. Zhu, Nat. Commun. \textbf{8}, 14956 (2017).

\bibitem{ref30}Y. Zhou \emph{et al.}, Nano Lett. \textbf{17}, 5508 (2017).

\bibitem{ref31}L. Ju, J. Shang, X. Tang, and L. Kou, J. Am. Chem. Soc.
\textbf{142}, 1492 (2020).

\bibitem{ref32}C. Gong, E. M. Kim, Y. Wang, G. Lee, and X. Zhang, Nat.
Commun. \textbf{10}, 2657 (2019).

\bibitem{ref33}G. Kresse and J. Furthmüller, Comput. Mater. Sci.
\textbf{6}, 15 (1996).

\bibitem{ref34}G. Kresse and J. Furthmüller, Phys. Rev. B \textbf{54},
11169 (1996).

\bibitem{ref35}Supplemental Material presents computational details;
additional electronic nad magnetic properties of TM decorated graphenen 
and their dependence on the FE polarization of the In$_2$Se$_3$ monolayer; 
variation of magnetism as a function of artificially increased interlayer 
distance; dependence of FE controlled magnetic moments on the choice of 
van der Waals corrections; PDOS shifts in the Cr@SVG/In$_{2}$Se$_{3}$
heterostructure; and also contains references \cite{a1,a2,a3,a4}.

\bibitem{a1}G. Kresse and J. Furthm\"uller, Comput. Mater. Sci. \textbf{6}, 15 (1996).
\bibitem{a2}G. Kresse and J. Furthm\"uller, Phys. Rev. B \textbf{54}, 11169 (1996).
\bibitem{a3}J. P. Perdew, K. Burke, and M. Ernzerhof, Phys. Rev. Lett. \textbf{77}, 3865 (1996).
\bibitem{a4}S. Grimme, J. Antony, S. Ehrlich, and H. Krieg, J. Chem. Phys. \textbf{132}, 154104 (2010).


\bibitem{ref36}M. L. Sun, Q. Q. Ren, Y. M. Zhao, J. P. Chou, J. Yu, and W.
C. Tang, Carbon \textbf{120}, 265 (2017).

\bibitem{ref37}V. M. Pereira, F. Guinea, J. M. Lopes dos Santos, N. M. R.
Peres, and A. H. Castro Neto, Phys. Rev. Lett. \textbf{96}, 036801 (2006).

\bibitem{ref38}M. M. Ugeda, I. Brihuega, F. Guinea, and J. M.
Gómez-Rodríguez, Phys. Rev. Lett. \textbf{104}, 096804 (2010).

\end{thebibliography}

\end{document}